\documentclass[10pt,sigconf,letterpaper]{acmart}
\setcopyright{none}

\renewcommand\footnotetextcopyrightpermission[1]{} 
\usepackage{gensymb}
\usepackage{booktabs}  
\usepackage{pbox}
\usepackage{subfigure}
\usepackage{epsfig,endnotes}
\usepackage{epstopdf}
\usepackage{times}
\usepackage{color}
\usepackage{graphicx}
\usepackage{xspace}
\usepackage{enumitem}
\usepackage{multirow}
\usepackage{tikz}
\usepackage[ruled,vlined,linesnumbered]{algorithm2e}

\newif\ifcomment
\commentfalse

\newif\ifwatermark
\watermarkfalse

\ifwatermark
      \usepackage{draftwatermark}
      \SetWatermarkScale{.7}
      \SetWatermarkText{\shortstack{In Submission:\\Do Not Distribute}}
      \SetWatermarkColor[rgb]{1,0.88,0.88}
\fi

\ifcomment
    \newcounter{MVNumberOfComments}
    \stepcounter{MVNumberOfComments}
    \newcommand{\mvnote}[1]{\textcolor{blue}{\small \bf [MV\#\arabic{MVNumberOfComments}\stepcounter{MVNumberOfComments}: #1]}}

    \newcommand{\NOTE}[1]
    {
      {\footnotesize\it
        \begin{center}
          \begin{tabular}{|c|}
           \hline
            \parbox{0.85\columnwidth}{
              \medskip
              #1
              \medskip} \\
            \hline
          \end{tabular}
        \end{center}
        }
    }
\else
    \newcommand\mvnote[1]{}
    \newcommand\NOTE[1]{}
\fi

\newcommand{\eg}{{e.g.,}\xspace}
\newcommand{\ie}{{\it i.e.,}\xspace}
\newcommand{\folder}{./fig}

\newcommand{\blab}{{BatteryLab}\xspace}
\newcommand{\dimtool}{{AttentionDim}\xspace}

\newcommand{\numandroid}{{15}\xspace}
\newcommand{\numbrowsers}{{15}\xspace}
\newcommand{\numdevices}{{6}\xspace}
\newcommand{\numusers}{{10}\xspace}
\newcommand{\numhours}{{500}\xspace}
\newcommand{\cappuccio}{{Cappuccino}\xspace}

\newcommand\co[1]{}

\begin{document}

\date{}

\title{On the Battery Consumption of Mobile Browsers}
\author{Matteo~Varvello$\dag$, Benjamin~Livshits$\dag\diamond$}
\affiliation{%
  \institution{$\dag$~Brave~Software, $\diamond$~Imperial~College~London}
}


\begin{abstract}
Mobile web browsing has recently surpassed desktop browsing both in term of popularity and traffic. Following its desktop counterpart, the mobile browsers ecosystem has been growing from few browsers (Chrome, Firefox, and Safari) to a plethora of browsers, each with unique characteristics (battery friendly, privacy preserving, lightweight, etc.). In this paper, we introduce a browser benchmarking pipeline for Android browsers encompassing automation, in-depth experimentation, and  result analysis. We  tested \numbrowsers Android browsers, using \textit{\cappuccio} a novel testing suite we built for third party Android applications. We perform a \emph{battery-centric} analysis of such browsers and show that: 1) popular browsers tend also to consume the most, 2)  \emph{adblocking} produces significant battery savings (between 20 and 40\% depending on the browser), and 3) \emph{dark mode} offers an extra 10\% battery savings on AMOLED screens. We exploit this observation to build \textit{\dimtool}, a screen dimming mechanism driven by browser events. Via integration with the Brave browser and 10 volunteers, we show potential battery savings up to 30\%, on both devices with AMOLED and LCD screens.
\end{abstract}

\maketitle

\section{Introduction}
\label{sec:intro}
When it comes to mobile apps, users are tied to the official app from the services they access. This is not the case for mobile browsers where plenty of options are currently available for both Android and iOS~\cite{browse-share}. While iOS browsers \textit{must} use Safari's rendering engine~\cite{webkit}, Android browsers are allowed more freedom although, in reality, most browsers rely on a common Chromium source base~\cite{chromium}. 


Such a competitive environment constantly stimulates the development of new browsers as well as new browser functionalities. In the last years, there has been a growing interest in reducing browsers (and apps in general) power consumption, motivated by the ever-increasing phone usage and app complexity. \emph{Adblocking}---either in the form of an addon~\cite{adblock, ublock} or directly integrated in the browser~\cite{brave,kiwi,opera}---is probably the most popular feature which has recently been connected with battery savings~\cite{greenspector_2018,heitmann2020towards}.  \emph{Dark theme}~\cite{dark-theme} is another feature which, originally introduced for eye strains, is now credited with high battery savings in presence of AMOLED screens which effectively turn pixels off when dark. The Yandex browser also offers a mysterious \emph{power saving mode}~\cite{yandex-power}. 

The goal of this work is to shed some light on the Android browser ecosystem. Our approach is clearly \emph{battery-centric}, but it also covers other metrics which directly impact battery usage, like CPU and bandwidth utilization. A strawman research approach to this problem consists in building a local testbed, \eg one Android device connected to a power meter, and writing automation code for a set of browsers and devices to be tested. 
Such approach does not offer \textit{reproducible} research, which is paramount to guarantee \textit{transparency} when commercial entities are involved. \textit{Scalability} is another issue given manual work can rapidly become overwhelming. 

Motivated by the above, we have built a generic browser testing suite --  which provides both fairness and transparency -- where \emph{human-generated} automation is plugged as needed. To do so, we have built \cappuccio the alter ego of the Espresso test recorder~\cite{espresso}. In the same way as Espresso can automatically generate testing \textit{code} from human input, \textit{\cappuccio} automatically generates \textit{automation} for third party apps. We integrated \cappuccio with Batterylab~\cite{batterylab2019} -- a research platform for battery measurements -- to realize a fully transparent and extensible browser testing suite. 

We used the above approach to benchmark the battery consumption (and more) of \numandroid Android browsers under different configurations, workloads, and devices. 
We find that: 1) popular browsers tend also to consume the most, 2) adblocking produces significant battery savings (between 20 and 40\% depending on the browser), 3) Yandex's power saving mode does not produce any extra battery saving, and 3) dark theme offers an extra 10\% of savings on AMOLED devices. The latter observation motivates us to build \textit{\dimtool}, a screen dimming mechanism driven by browser events. Via integration with a commercial browser and 10 volunteers, we show that \dimtool can further reduce battery consumption up to 30\%, independently of the device's screen technology.

\section{Benchmarking Pipeline}
\label{sec:meth}
The underlying goal of this paper is to benchmark the energy consumption of multiple Android browsers. 
A \textit{strawman} approach to this research question requires: 1) building a \textit{local} testbed composed of an Android device and a power meter~\cite{buiMOBICOM15, caoPOMAC17, ravenMOBICOM17,thiagarajanWWW12}, 2) write code to automate each browser, \eg how to launch and instrument, 3) write code to instrument the device and the power meter, \eg collect performance metrics and minimize experimental noise.

Such strawman approach does not provide \textit{reproducible} research, which we argue is a necessity when commercial products are at play. Further, it does not \emph{scale} given that automation code needs to be written per browser and device. In fact, while some operations are common across browsers, \eg launch and open a URL, others are browser-specific, rarely exported as \textit{flags} or unusable on regular Android devices~\cite{chrome_engineer_mode}. 

Fortunately, the research community has recently released \blab~\cite{batterylab2019}, a testbed which largely simplifies battery measurements. In short, \blab consists of a set of remote devices connected to power meters where \textit{experimenters} can run ad-hoc experiments.  \blab also offers simple APIs to collect fine-grained battery readings along with other metrics like CPU and bandwidth usage. This testbed not only eliminates two of the three limitations of the strawman approach above, but further fosters our \textit{transparency} goal. Accordingly, we are left with the need of building a \textit{generic} browser testing pipeline, which we describe in the upcoming subsections. 


\subsection{Human Driven Automation}
\label{sec:meth:automation}
\begin{figure}[t]
    \centering
    \psfig{figure=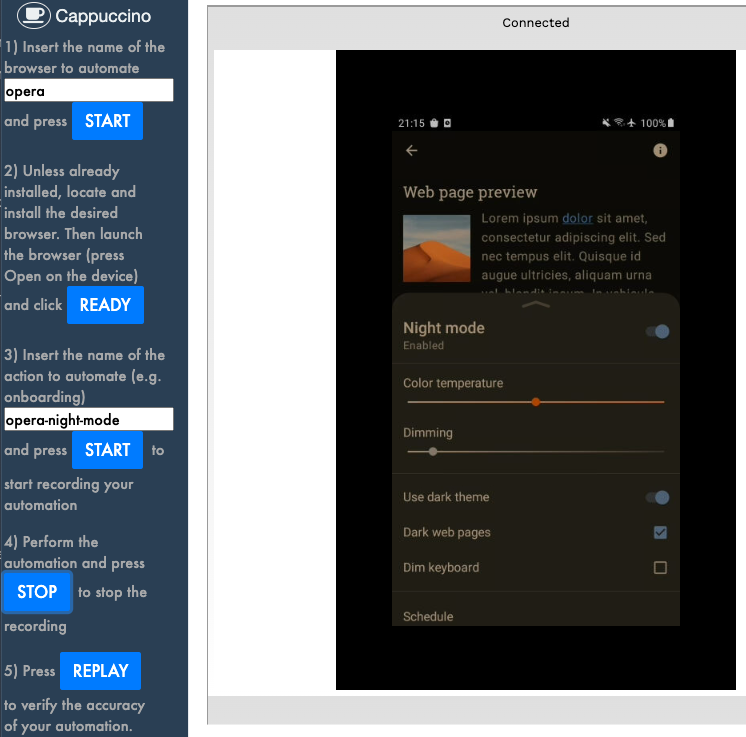, width=\columnwidth}
    \caption{\cappuccio's GUI. On the left, the user follows instructions to record an automation. On the right, a virtual device is shown. The example refers to the generation of Opera's \textit{night-mode} selection.}
    \vspace{-0.1in}
    \label{fig:webtest-cdf-human}
    \vspace{-0.1in}
\end{figure}

Today, testing third party Android apps (\ie lack of source code) requires an experimenter to write his/her own automation by manually interacting with a device while recording the actions executed. Such actions can then be translated into Android Debugging Bride (ADB~\cite{adb}) commands to generate \textit{automation scripts}. Better tools are instead available for developers who have access to the code to be tested. For example, Espresso test recorder~\cite{espresso} automatically generates testing code from natural interaction with an Android emulator. In the following we describe \textit{\cappuccio}, the equivalent of Espresso for third party apps. 

The intuition behind \cappuccio derives from the work in~\cite{almeida2018chimp}, where the authors crowdsource human input by streaming (emulated) Android devices in the browser. In the same spirit, \cappuccio streams a real (or emulated) device in the browser where a custom version of noVNC records the user input and maps it to \textit{generic} ADB commands, with the goal to build automated scripts. We say these commands are generic since they are expressed as ratios of screen resolutions -- accounting, for example, for on screen toolbars as in Samsung devices -- so to be reused across several devices. 

Figure~\ref{fig:webtest-cdf-human} shows the workflow of \cappuccio's human input collection. First, an experimenter provides information about the app to be tested, and proceed with installation and launch. This allows \cappuccio to \textit{learn} how to launch the newly installed app, \ie package and activity name to be used.
Next, the experimenter start generating \textit{automations}. Some predefined \textit{automation\_labels} are offered (\eg onboarding) and customs are possible to (\eg enableAdblocking). Start and stop recording buttons are used to inform \cappuccio on when to start and stop recording human inputs. A \textit{replay} button is also available to execute the automation script just derived from the human input. In this way, an experimenter can evaluate the accuracy of the automation generated and decide whether to save it or not. 

Thanks to the \blab team, we integrated \cappuccio with \blab. This means that app/browser automation can be saved at \blab's access server under pair $<$browser, automation\_label$>$ and easily accessed during browser testing (see next section). All automations used in this paper were produced by \cappuccio. A word of caution is needed though since browser automations are simple as mostly require clicks and, at most, few scrolls. More complex app automations can be hard to replay, especially due to divergent behavior between devices. Further evaluating \cappuccio to a broader set of mobile apps is part of our future work. 

\subsection{Browser Testing Pipeline}
\label{sec:meth:testing}
Algorithm~\ref{alg:browser-automation} shows the pseudocode of a generic \textit{job} for browser testing. Such job targets \blab and as such it relies on its APIs, \eg device preparation and battery measurements. However, it can  be extended to run on other device farm solutions granted that they allow ADB~\cite{adb} access to their devices,  like for instance Samsung's Remote Test Lab~\cite{samsung}. In the following, we capitalize all calls to \blab's API. 

Our generic browser testing job expects as input the id of the device where to run (\textit{device}), the list of browsers to be tested (\textit{browser\_list}), JSON containing the desired automation (\textit{automation\_dict}), and JSON containing the websites to be tested and how (\textit{workload\_dict}), \eg load in a new tab, time spent on site, webpage interaction strategy. The browser testing workflow consists of four phases: \emph{device setup}, \emph{browser setup}, \emph{data collection}, and \emph{testing}. 

\vspace{0.05in}
\noindent \textbf{Device Setup} -- The device under test is configured to minimize noise on the measurements. For instance, background processes and app notifications are disabled (\texttt{DEVICE\_SETUP}, L1 in Algorithm~\ref{alg:browser-automation}).

\vspace{0.05in}
\noindent \textbf{Browser Setup} -- Before a test, a browser might need to be installed (L4). By default, the \texttt{INSTALL} API relies on the PlayStore and thus installs the most recent version of an app. For custom testings, a URL pointing to the \texttt{.apk} to be tested can also be provided. Next, the browser is setup with a clean \textit{profile}, \ie its cache is emptied and local data like configuration, cookies, and history are erased. This step (L5) is equivalent across browsers since it relies on the OS to \textit{clean} a target app/package.
Next, we deal with a browser's \textit{onboarding} process (L6), a common operation where the user is asked to customize the browser, \eg by choosing a search engine or turning adblocking on/off. This step differs across browsers and it is thus the first step where we rely on the human-driven automation described above. This is the first step where the browser can be customized for a specific setting to be tested by the experimenter. More \textit{human-driven} settings are setup with the for loop in L7-L8. 

\vspace{0.05in}
\noindent \textbf{Data Collection} -- Once both phone and browser are configured for a test, we enter the ``data collection'' phase where fine-grained battery measurements (1,500 current/voltage samples per second), CPU and memory usage (5 seconds frequency, via \texttt{/proc/stat}), bandwidth consumption (via \texttt{/proc/net/}) are collected. This phases only starts when the CPU load returns to common rest values (between 0 and 5\% for more than 15 seconds) after the CPU spikes caused by device and browser preparation.

\vspace{0.05in}
\noindent \textbf{Testing} -- 
For each page, the browser launches it, wait for a certain amount of time, and interact (or not) with the page executing several scrolls up and down. Pages and load details are described in $workload\_dict$, either as pure ADB commands on \cappuccio automations. 

\begin{algorithm}[t]
    \DontPrintSemicolon
    \KwIn{Device identifier $device$, Browser list $browser\_list$, Automation JSON  $automation\_dict$, Workload JSON $workload\_dict$}
    \KwOut{JSON file with performance metrics}
    \SetAlgoLined
    \SetKwComment{tcp}{$/\!\!/$ }{}
    \SetCommentSty{normal}
    \BlankLine
    {\it device\_status} $\leftarrow$ {\sc device\_setup}({\it device})\\    
    \For(){$browser \in  $browser\_list}{
        {\it dict} $\leftarrow$ {\sc  automation\_dict[browser]}\\
        {\sc install}({\it browser})\\
        {\sc clean}({\it browser})\\
        {\sc onboarding}({\it browser}, {\it device}, dict["onboarding"])\\
        \For(){$setting \in $dict[browser]["settings"]}{
            {\sc setup}({\it device}, {\it browser}, {\it setting})\\
            }
        {\sc data\_collection}({\it device})\\ 
        {\sc run\_test}({\it device}, {\it browser}, {\it url\_list}, {\it workload})\\
    }
    {\it device\_status} $\leftarrow$ {\sc CLEANUP}({\it device})   
\caption{Pseudo-code for browser testing.}
\label{alg:browser-automation}
\end{algorithm}
\vspace{0.2in}

\begin{figure*}[thb]
   \centering
   \subfigure[Battery discharge (mAh).]{\psfig{figure=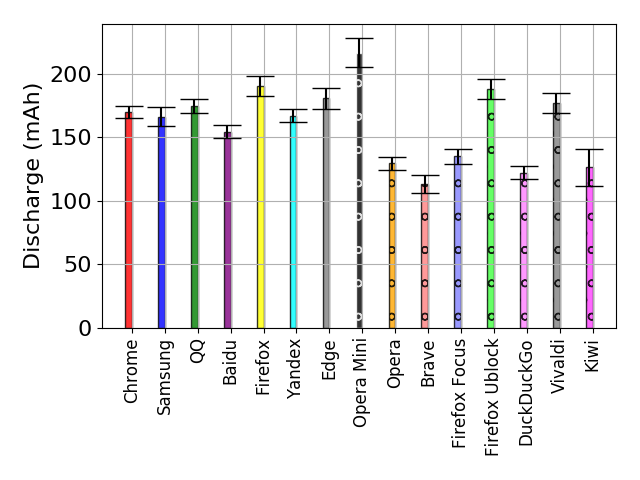, width=2.25in}\label{fig:res:android_discharge}}
   \subfigure[Bandwidth consumption (MBytes)]{\psfig{figure=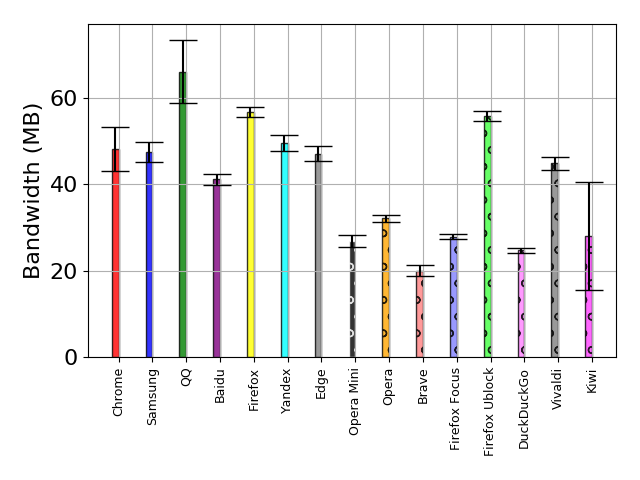, width=2.25in}  \label{fig:res:android_bdw}}           
   \subfigure[CDF of CPU utilization during a single run.]{\psfig{figure=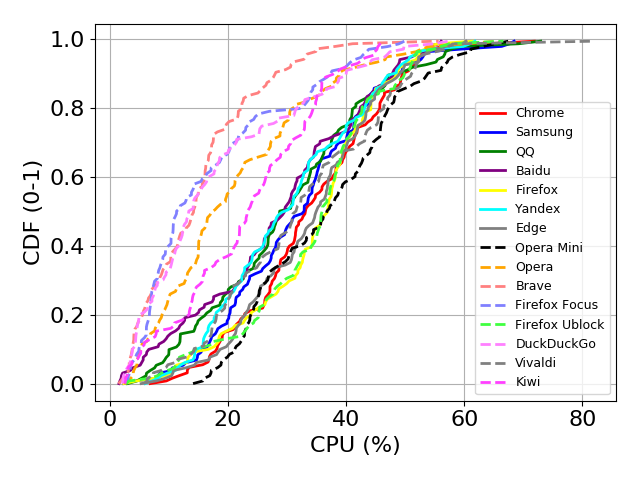, width=2.25in} \label{fig:res:android_cpu}}      
   \caption{Performance evaluation of \numbrowsers Android browser ; News workload; J7DUO}
   \label{fig:results_android}
\end{figure*}


\section{Browsers Evaluation}
\label{sec:res}
This section offers an empirical evaluation of the Android browser ecosystem. We start by describing browsers, workloads, and devices under tests, along with the rationale beyond their selection. Next, we report on the benchmarking results. 

Our rationale for browser selection is threefold. First, we target \emph{popular} browsers which are indeed used in the wild. Second, we target \emph{adblocking} browsers -- either native or enhanced with adblocking addons -- because of the potential energy benefits associated with adblocking~\cite{greenspector_2018,heitmann2020towards}. Finally, we target browsers which advertise energy saving capabilities.  Based on this strategy, we have selected \numandroid browsers to be tested (see Table~\ref{tab:summ_browser}, Appendix~\ref{sec:appendix}). 

We call \textit{workload} the content of the  $workload\_dict$ JSON file (see Section~\ref{sec:meth:testing}) describing \textit{which} websites should be tested and \textit{how}. According to~\cite{num_tabs}, most users keep the number of open tabs between 1 and 10. Accordingly, we opted to open testing webpages sequentially as a new tab. Each page is requested for $T$ seconds (empirically estimated to 10 seconds) to allow full page loads. Note that waiting for \texttt{onload} only work for some of the Chromium-based browsers, and would cause uneven experiment durations. Next, we simulate multiple user interactions by scrolling the page down $N$ times and then up $N/2$ times, for 30 seconds. With respect to the pages to be tested we pick 10 popular news websites around the world (\textit{news} workload) as well as 10 (hard to find) ads-free websites (\textit{ads-free} workload). The rationale of our selection is that these two workloads are realistic representation of, respectively,  the best and worst case scenarios for adblocking browsers. The full list of websites selected can be found in Table~\ref{tab:workload}) (Appendix \ref{sec:appendix}).

From \blab, we use two 2018 devices: a Samsung J7 Duo (J7DUO) and a Samsung Galaxy J3 (SMJ337A). The main difference between the two is their screen technology (AMOLED for the J7DUO and LCD for SMJ337) which is expected to offer significant differences when measuring the battery savings associated with dark mode. With respect to their hardware, the J7DUO is more powerful, with twice as many cores (octa vs quad-core) and RAM (4 vs 2GB). 

\subsection{Popular or Adblocking?}
\label{sec:meth:selection}
Figure~\ref{fig:res:android_discharge} summarizes the performance evaluation (battery discharge, bandwidth consumption, and CPU utilization) of all browsers under test with default configuration, while considering news workload and J7DUO. Barplots report, for each metric, the average with errorbars for standard deviation (values computed over 5 runs). Given the CPU consumption evolves over time, we instead report one representative Cumulative Distribution Function (CDF) per browser (see Figure~\ref{fig:res:android_cpu}). We use circle markers (barplots) and dashed lines (CDFs) to highlight adblocking browsers. 

Figure~\ref{fig:res:android_discharge} shows that the most popular browsers are quite similar with respect to battery consumption, with Baidu leading the pack with minimum consumption at 150mAh. Most adblocking browsers are more power efficient, with the exception of Vivaldi and Firefox equipped with the Ublock plugin, and Opera Mini which shows a staggering 225mAh during our test. Among adblocking browsers, Brave consumes the least followed by Opera and Kiwi which suffers from higher standard deviation than most browsers. Figure~\ref{fig:res:android_bdw} justifies this result showing that adblocking browsers can save tens of MBytes by non downloading ads. Not all ad-blockers are equal though, most notably Firefox uBlock seems quite relaxed and Kiwi suffers, again, from quite variable results -- potentially due to a less mature development. 

To explain the strange case of Opera Mini, consuming little bandwidth but high battery, we resort to Figure~\ref{fig:res:android_cpu}, which shows the CDF of the CPU consumption during our tests, per browser. Opera Mini shows the highest CPU consumption, median of 35\% with peaks up to 70\%,  twice as much as browsers like Brave and Firefox focus, with a median of 15\%.

\begin{figure}[b]
    \centering
    \psfig{figure=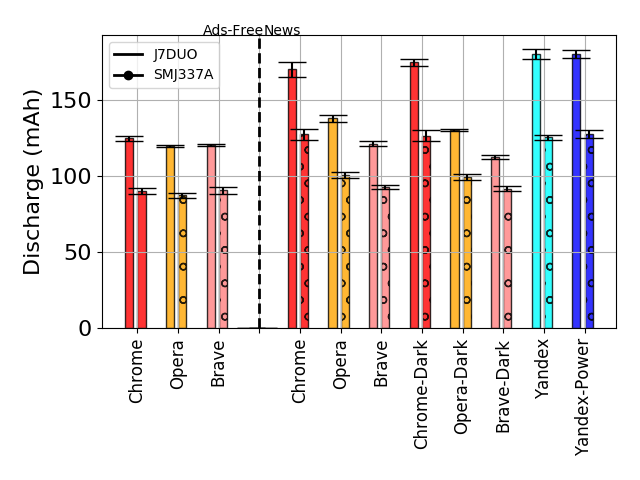, width=\columnwidth}
    \caption{Battery discharge of adblocking, dark mode, and Yandex power saving. Ads-Free and News work are on the left/right of the dashed line. Results refer to both J7DUO and SMJ337A (circle markers).}
    \label{fig:res:powermode}
    \vspace{-0.15in}
\end{figure}

\subsection{Power Saving} 
\label{sec:power:adhoc}
The previous subsection shows that \textit{adblocking} results in significant power savings. We here generalize the analysis with respect to other techniques, namely \textit{dark mode} -- which can potentially save battery by allowing to turn pixels off on AMOLED screens -- and Yandex power saving mode, to the best of our knowledge, the only explicit power saving feature offered by an Android browser. In this analysis we also introduce the \textit{ads-free} workload, to estimate the potential cost of adblocking in absence of ads. Further, we introduce the second device (SMJ337) which is still equipped with an LCD screen, \ie dark pixels are indeed not turned off. To reduce the measurement space, we select a mix of popular and best performing browsers with different levels of adblocking: Chrome, Opera, and Brave. 

Figure~\ref{fig:res:android_discharge} shows the energy consumption across browsers, devices, and workloads. The figure shows that in absence of ads (left of the dashed line) browsers have very similar battery consumption, irrespective of the devices. We also notice that less powerful SMJ337 also consumes $\sim$20\% less than the J7DUO. This could be due to many things, such as the bigger screen and overall more advanced hardware to be powered. 

Next we focus on the \textit{news} workload. Regardless of the device, the trend is the the same as the one observed before with the most aggressive adblocking browser (Brave) bringing the highest battery savings. In addition, dark mode offers about 11-13\% extra savings for Opera and Brave, respectively. For Chrome-Dark we do not measure additional savings; the ``*'' indicate that for Chrome we selected the basic dark mode settings -- \ie the one available via GUI without accessing \textit{chrome://flags} -- which does not darken the page but only the browser's GUI. We purposely selected this mode to measure potential benefits from this feature, which are negligible in our tests. As expected, dark mode only brings benefits to the J7DUO since SMJ337 does not mount an AMOLED screen. 
Last but not lest we did not observe any difference when activating Yandex's power saving mode, despite the 9\% battery saving advertised. 

\newcommand{\numsers}{{10}\xspace}
\newcommand{\numdays}{{30}\xspace}

\begin{table}
\scriptsize
\centering
\begin{tabular}{|c|c|c|c|c|c}
\hline 
\bf Brightness & \bf Scenario  & \bf \begin{tabular}{@{}c@{}} Current \\ (mA, median)\end{tabular} & \bf \begin{tabular}{@{}c@{}} Savings \\ (Aggressive)\end{tabular}  & \bf \begin{tabular}{@{}c@{}} Savings \\ (Conservative)\end{tabular} \\
\hline
\hline
\bf 0   & -              &  145/108   & 0/0\%    & 0/0\%     \\ \hline
\bf 50  & Indoor         &  189/130   & 23/17\%  & 23/17\%   \\ \hline
\bf 100 & Indoor         &  239/157   & 39/31\%  & 39/31\%   \\ \hline
\bf 150 & Cloud Outdoor  &  299/201   & 51/46\%  & 28/28\%   \\ \hline
\bf 200 & Outdoor        &  379/243   & 61/55\%  & 21/17\%   \\ \hline
\bf 250 & Sunny Outdoor  &  417/247   & 65/56\%  & 28/17\%   \\ \hline
\end{tabular}
\caption{Screen brightness reduction power savings (J7DUO/SMJ337A)}
\label{tab:dimming}
\vspace{-0.2in}
\end{table}

\section{\dimtool}

While browsing a user ``waste'' quite some time, \eg when typing a URL or while a webpage is loading. For example, under \textit{bad} network conditions a user can spend tens of seconds waiting for a webpage to load, and eventually give up with no content displayed. Our intuition is to minimize the screen power consumption during these wasted times.  We thus propose ``\dimtool'', a screen dimming strategy which leverages the browser state, \eg loading versus content ready, to define screen brightness.

We motivate this idea by investigating the \textit{potential} savings deriving from screen dimming. Table~\ref{tab:dimming} shows, for several increasing brightness values in Android (\ie 0-250 range) and corresponding scenarios where they apply, the median current (mA) measured on both J7DUO and the SMJ337A during one minute displaying a default Android desktop theme. The table further extrapolates the potential battery savings coming from \emph{full} screen dimming, \ie dropping the screen brightness to zero, as well as the more \textit{conservative} strategy we will detail below. This experiment shows that, even with a conservative strategy, screen dimming offers potential savings between 17 and 40\%, on both AMOLED and LCD-equipped devices. 

The above savings highly depend on actual device usage, \eg slow mobile networks offer more opportunities for savings. Accordingly, rather than focusing on in-lab experiments, we have directly integrated \dimtool in the Brave browser (Android) and performed experiments in the wild. We picked Brave since it resulted one of the ``greener'' browser from the previous analysis. Nevertheless, the code is generic and can be used by any Chromium-based browser. Implementation and evaluation details are reported in the following. 

\begin{figure*}[t]
   \centering
   \subfigure[CDF of fraction of time spent dimming.]{\psfig{figure=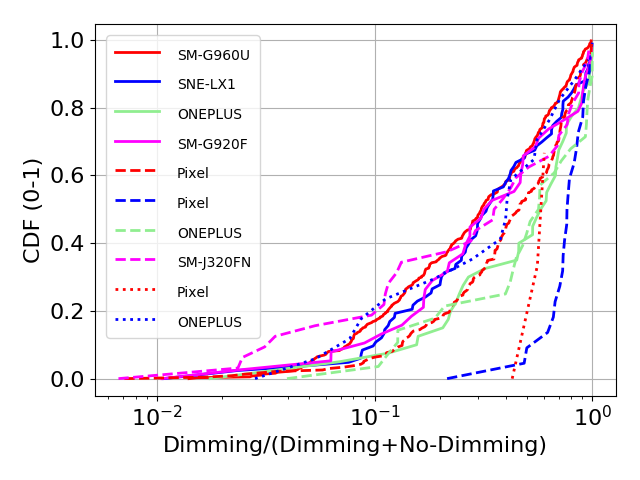, width=2.1in} \label{fig:res:dimming:1}}
   \subfigure[CDF of screen brightness.]{\psfig{figure=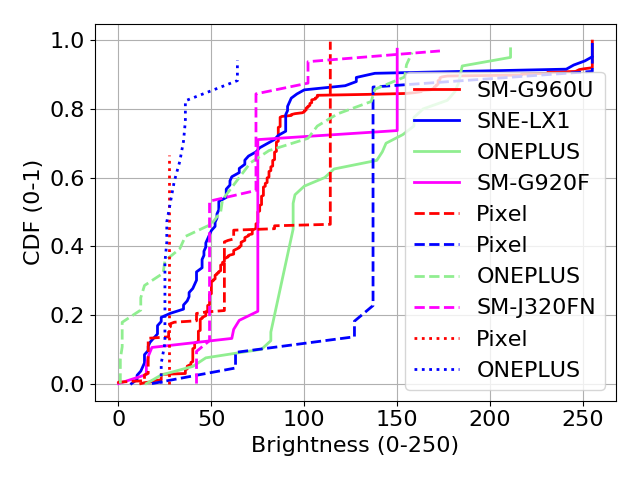, width=2.1in} \label{fig:res:dimming:2}}
   \subfigure[CDF of estimated energy savings.]{\psfig{figure=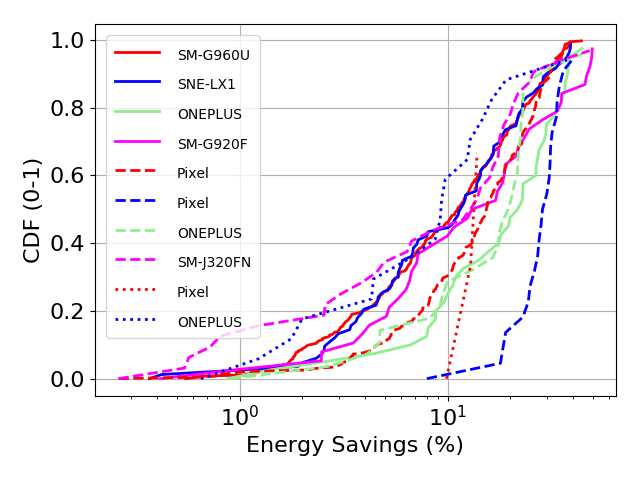, width=2.1in} \label{fig:res:dimming:3}} 
   \caption{Performance evaluation of \dimtool in the wild.}
   \vspace{-0.1in}
   \label{fig:results:dimming:wild}
 \end{figure*}

\subsection{Design and Implementation}
The idea behind \dimtool is to use \texttt{onLoad()}, a browser event which signals when a page is loaded, as an approximation of user \textit{attention} which requires regular screen brightness. We have identified three events \textit{where} user attention is lowered: 1) URL typing, 2) menu settings, 3) webpage loading. Personal preferences are at play, but \dimtool is thought for the user who is willing to sacrifice a bit of his/her user experience for longer battery duration. Other ``events'' are possible, \eg video buffering, but requires more complex browser modifications and were thus left as future work. 

\dimtool consists of a  module  which controls the screen brightness from the browser. This module currently sits in \texttt{ChromeTabbedActivity}, \ie it can be adopted by all Chromium-based browsers, and it is triggered by the above events to \textit{dim} the screen and then \textit{restore} the  brightness when the event completes. When dimming is triggered, this module detects whether the user is using \textit{auto} or \textit{manual} brightness so that it can: 1) manually restore the previous brightness value when the event completes, 2) reactivate auto dimming and let the OS decide the brightness value to be restored. 

We experimented with several dimming strategies and then settle for the following one based on feedback received from our volunteers. 
When the original screen brightness is \textit{low} (\ie <= 100) we opt for an aggressive strategy, \ie we lower the screen brightness $B$ to zero. For \textit{mid} brightness values (\eg 150) we set $B$ to half of the original value (50 and 75). We instead use a fixed $B=150$ for \textit{high} values, since outdoor and sunny conditions are quite challenging and we need to prevent leaving users in the dark. Last but not least, we implemented a setting option and a simple GUI to allow users to deactivate \dimtool when in trouble and to get a sense of the battery savings provided. 


\subsection{Evaluation}

We recruited \numsers Android users who installed our modified version of Brave for up to \numdays consecutive days totaling about \numhours hours of browsing -- we urged our volunteers to use the browser as normal. The test spans \numdevices devices
since multiple volunteers shared the same device model. 


Figure~\ref{fig:res:dimming:1} shows the CDF of the fraction of time spent dimming, per device. The CDF is calculated using the beginning of a dimming event as both the start time of such event and the end time of the previous non-dimming event. Start/end timers are also triggered whenever the user closes or (re)launch the browser. As expected, the amount dimming is very much user and time-dependent, meaning that some users experience a higher amount of dimming as well as the dimming duration spans a broad distribution. Generally speaking, very short dimming events (\eg lower than 10\%) are rare. Across users, the median dimming event lasts between a minimum of 30 and a maximum of 70\% of the time. 

Figure~\ref{fig:res:dimming:2} shows the CDF (per device) of the screen brightness hen \dimtool operated. Most brightness values reported are smaller than 100 (indoor usage). One of the Pixel devices is an exception since most values reported were quite high, either because of outdoor or manually set. It has to be noted that this device was also only used for a limited amount of time, as the sharp CDF suggests.  Finally, we combine the information from Figure~\ref{fig:res:dimming:1},~\ref{fig:res:dimming:2}, and Table~\ref{tab:dimming} to estimate actual battery savings. 
The figure shows encouraging battery savings of about 20-30\%, which means potentially extending the battery life by up to one hour.

\section{Related Work}
Browser benchmarking is a largely discussed topic in the greater ``web community'', while less attention was dedicated by the research community. In the following we report on two research papers and one blog which, to the best of our knowledge, share some similarities with our work. 

Nielson et al.~\cite{nielson2008benchmarking} benchmark  four  popular browsers (Firefox, IE, Opera, and Safari) at the time of this work (2008). Their results show substantial  differences  among browsers  across  the  range  of  tests  considered,  particularly  in rendering  speed and  JavaScript  string operation  performance. Our work is similar in spirit to~\cite{nielson2008benchmarking} but differs under many aspects. First, the browsing ecosystem largely changed in the last 10 years, \eg the prevalence of Android and mobile browsing which was not the case at that time. Second, our work also aims at building an automated and scalable testing suite that offers both \textit{transparency} and \textit{reproducibility}. 

Greenspector~\cite{greenspector_2018} -- a startup offering software-based battery measurements -- has recently ranked Android browsers based on their energy consumption (and other metrics). We perform a similar evaluation but relying on highly \textit{accurate} hardware-based measurements versus (proprietary) software-based measurements. Further, we test more workloads, websites, and features (such as dark mode, for instance). Nevertheless, our testing suite was designed to foster extensible and reproducible research in browser performance. 

Last but not least,~\cite{heitmann2020towards} analyzes the power consumption of the Brave browser, with respect to adblocking functionalities, over the Odroid-XU3 development board~\cite{odroid}. We also report numbers with respect to Brave, but in the context of two real Android devices. In our tests, we could not verify the reported results with respect to the extra cost of adblocking when ads are indeed missing (see Figure~\ref{fig:res:powermode}). 

\section{Conclusion}
This paper has investigated the battery consumption of \numbrowsers Android browsers, 3 top performing browsers in \textit{dark mode}, Yandex power saving feature, and \textit{\dimtool}, a novel screen dimming mechanism driven by browser events like \texttt{onLoad}. Given the scale of our measurements, we have also built a fully automated testing suite. For browser automation, we have built \textit{\cappuccio}, the first record and replay tool for third party Android apps. For browser testing, we integrated with  \blab a research platform which allows remote hardware-based battery measurements on a few Android devices. Our results show that \textit{adblocking} offers significant battery savings (around 30\%) which can be further enhanced via \textit{dark mode} (extra 10\%). Yandex power saving feature resulted more a marketing stunt than a beneficial solution. Finally,  we integrated \dimtool in the Brave browser -- one of the top performing browsers -- and run an experiment in the wild involving \numusers users and up to \numhours hours of real browsing. Our results show that \dimtool reduced, on average, the battery consumption of our volunteers by about 20-30\%.

\newpage

\bibliographystyle{acm}
\bibliography{biblio}

\appendix

\begin{table}[th!]
\scriptsize
\centering
\begin{tabular}{rccc}
\hline

           {\bf Browser}  & {\bf Version} & {\bf Chrome/Firefox Vrs} & {\bf Popularity}\\
\hline
{\bf Chrome}            & 81.0.4044.138    & 81.0.4044.138 & 89\%     \\
{\bf QQ}                & 10.3.0.6730      & 66.0.3359.126 & 2.8\%   \\
{\bf Samsung Browser}   & 11.1.2.2         & 75.0.3770.143 & 2.4\%   \\
{\bf Opera Mini}        & 47.2.2254.147957 & 81.0.4044.138 & 1.15\%  \\
{\bf Baidu}             & 4.14.0.30        & 81.0.4044.138 & 0.79\%  \\
{\bf Opera}             & 57.2.2830.52651  & 80.0.3987.162 & 0.64\%   \\
{\bf Firefox}           & 68.8.0           & Gecko/68.0 -  & 0.44\%   \\
{\bf Yandex}            & 20.3.4.98        & 80.0.3987.132 & >100M\%      \\
{\bf Edge}              & 45.03.4.4955     & 77.0.3865.116 & >10M          \\
\hline
{\bf Brave}            & 1.7.102           & 81.0.4044.122 & >10M     \\
{\bf Firefox-focus}    & 8.2.0             & Gecko/68.0    & >5M              \\
{\bf Firefox uBlock}   & 68.8.0            & Gecko/68.0    & >5M    \\
{\bf DuckDuckgGo}      & 5.52.6            & 81.0.4044.138 & >10M   \\
{\bf Vivaldi}          & 3.0.1885.43       & 81.0.4044.142 & >100K   \\
{\bf Kiwi}             & Quadea            & 77.0.3865.92  & >1M     \\
\hline
\end{tabular}
\caption{Android browsers selected for performance evaluation. Versions refer to the most recent version available at the time of testing (May 2020).}
\label{tab:summ_browser}
\end{table}

\vspace{0.2in}
\section{Browser and Workload Details}
\label{sec:appendix}

Table~\ref{tab:summ_browser} summarizes the name, version, and underlying engine (\ie Chrome or Firefox) of the \numbrowsers browsers we have tested in this paper. The popularity column was derived from data reported in~\cite{browser_share} on May 2020. For browsers not contemplated in this report, we use the most recent number of downloads from the Google Play Store. 

Table~\ref{tab:workload} shows the sequence of websites tested under the news (see Figure~\ref{fig:results_android}) and ads-free (see Figure~\ref{fig:res:powermode}, left of dashed line) workloads. Websites are each  loaded sequentially in a new tab. To find ads-free websites, we crawled Alexa top 1,000 websites and matched the content they served against the easy privacy list~\cite{easyprivacy}.  We then manually tested each website that did not trigger the easy privacy list in Brave in order to verify functionalities, \eg response times within 10 seconds, and lack of ads. We finally selected popular websites meeting the above criteria.


\begin{table}[t]
\small
\centering
\begin{tabular}{rccc}
\hline
    {\bf Ads-Free}                      & {\bf News}  \\
    \hline
    https://qwant.com                   &   https://theblaze.com       \\
    https://panda.tv                    &  https://thedailybeast.com  \\
    https://mega.nz                     &  https://independent.co.uk  \\
    https://i-register.in               &  https://nypost.com         \\
    https://hamariweb.com               &  https://salon.com          \\
    https://www.ipko.pl                 &  https://sfgate.com         \\
    https://bankmellat.ir               &  https://latimes.com        \\
    https://jw.org                      &  https://cnn.com            \\
    https://www.sarzamindownload.com/   &  https://mirror.co.uk       \\
    https://www.icloud.com/             &  https://cnet.com           \\
    \hline
\end{tabular}
\caption{Workload description (news and ads-free)}
\label{tab:workload}
\end{table}

\end{document}